\documentclass[epj]{svjour}
\usepackage{graphics}
\begin{document}
%
\title 
{
On the Sensitivity of L/E Analysis of Super-Kamiokande Atmospheric Neutrino Data to Neutrino Oscillation Part~2
}
\subtitle
{Four Possible L/E Analyses for the Maximum Oscillation by the Numerical Computer Experiment}

\author{E. Konishi\inst{1}, Y. Minorikawa\inst{2}, V.I. Galkin\inst{3},
M. Ishiwata\inst{4}, I. Nakamura\inst{4}, N. Takahashi\inst{1}, M. Kato\inst{5}  \and A. Misaki\inst{6}\inst{7}    
}
\institute{
Graduate School of Science and Technology, Hirosaki University, Hirosaki, 036-8561, Japan    
\and Department of Science, School of Science and Engineering, Kinki University, Higashi-Osaka, 577-8502, Japan
\and Department of Physics, Moscow State University, Moscow, 119992, Russia
\and Department of Physics, Saitama University, Saitama, 338-8570, Japan
\and Kyowa Interface Science Co.,Ltd., Saitama, 351-0033, Japan
\and Inovative Research Organization, Saitama University, Saitama,
 338-8570, Japan
\and Research Institute for Science and Engineering, Waseda University, Tokyo, 169-0092, Japan 
\\\email{konish@si.hirosaki-u.ac.jp}
}

\abstract {
In the previous paper (Part~1), we have verified that 
{\it the SK assumption on the direction} does not hold in 
the analysis of neutrino events occurred inside the SK detector.  
Based on the correlation between 
$L_{\nu}$ and $L_{\mu}$ 
(Figures~12 and 13 in Part~1) 
and the correlation between $E_{\nu}$ and $E_{\mu}$ 
(Figure 14 in Part~1), 
we have made four possible $L/E$ analyses, namely
$L_{\nu}/E_{\nu}$, $L_{\nu}/E_{\mu}$, $L_{\mu}/E_{\nu}$
 and $L_{\mu}/E_{\mu}$.
 Among four kinds of $L/E$ analyses, we have shown that only
 $L_{\nu}/E_{\nu}$ analysis can give the signature of maximum 
oscillations clearly, not only the first maximum oscillation 
but also the second and third maximum oscillation,
while the $L_{\mu}/E_{\mu}$ analysis which are really done 
 by Super-Kamiokande Collaboration cannot give the maximum 
oscillation at all.
It is thus concluded from those results that the experiments 
with the use of the cosmic-ray beam for neutrino oscillation,
 such as Super-Kamiokande type experiment, cannot find the 
maximum oscillation from $L/E$ analysis, because the incident
neutrino cannot be observed due to its neutrality.
Therefore, we would suggest Super-Kamiokande Collaboration 
to re-analyze the zenith angle distribution of the neutrino events 
which occur inside the detector carefully,
 because $L_{\nu}$ and $L_{\mu}$ are alternative expressions of 
the cosine of the zenith angle for the incident neutrino 
and that for the emitted muon, respectively. 
}
\authorrunning{E.Konishi et. al.,}
\titlerunning{On the Sensitivity of L/E Analysis of SK Neutrino Oscillation}
\maketitle
%

\section{Introduction}


 In Figures~12 and 13 of the preceding paper\cite{Konishi2},
 we have shown that {\it the SK assumption on the direction}
 that the directions of the incident neutrinos are the same as those 
of the emitted muons does not hold.
 Also, in Figure~14 of the same paper, we have shown that the 
energies of the incident neutrinos cannot be determined from the 
those of the emitted muons, uniquely.
 However, the discrepancies between two variables in 
Figures 12 and 13 are distinctively large compared with those in 
Figure~14. 
Therefore, non-holding of {\it the SK assumption on the direction} 
plays an essential role in the $L/E$ analysis for finding the 
maximum oscillation (oscillation pattern in neutrino oscillation).

The survival probability of a given flavor is given in Eq.(1),
 in the case of Super-Kamiokande Collaboration. 
The variables for the $L/E$ analysis are $L_{\nu}$ and $E_{\nu}$,
 where $L_{\nu}$ denotes the flight length for the incident 
neutrino between the generation point of the incident neutrino
 and the interaction point of the neutrino concerned in the detector,
 and $E_{\nu}$ is the energy of the incident neutrino.

\begin{figure}
\begin{center}
\resizebox{0.45\textwidth}{!}{%
  \includegraphics{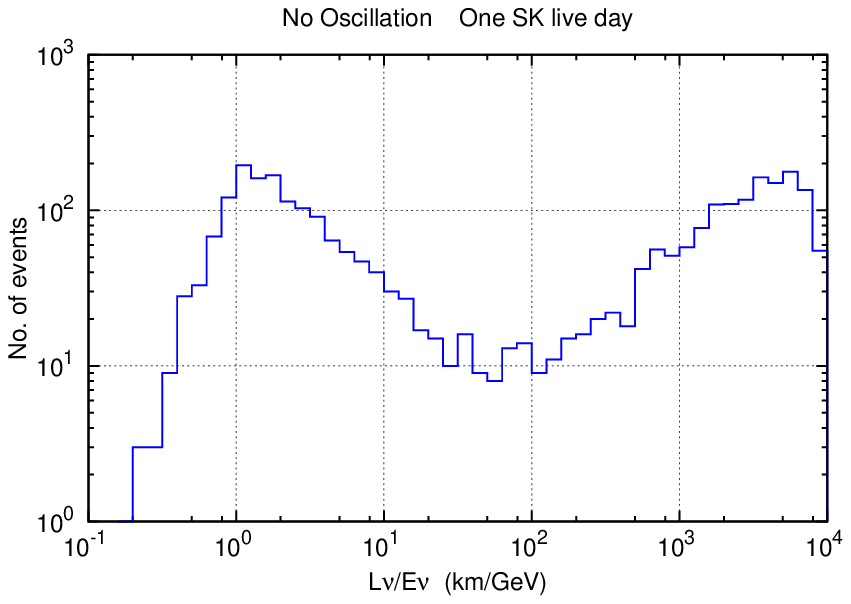}
}
\caption{$L_{\nu}/E_{\nu}$ distribution without oscillation
for 1489.2 live days (one SK live day).}
\label{figJ015}       
\resizebox{0.45\textwidth}{!}{%
  \includegraphics{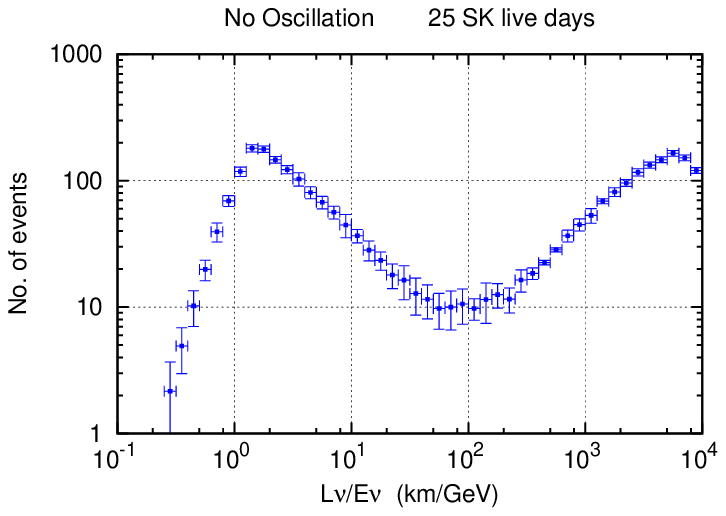}
}
\caption{$L_{\nu}/E_{\nu}$ distribution without oscillation
for 37230 live days (25 SK live days).}
\label{figJ016}       
\end{center}
\end{figure}
\begin{figure}
\begin{center}
\resizebox{0.4\textwidth}{!}{%
  \includegraphics{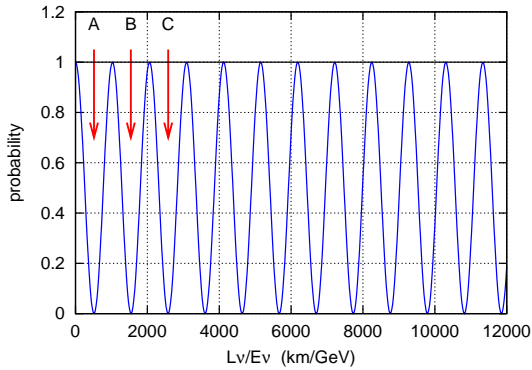}
}
\caption{Survival probability of
$P(\nu_{\mu} \rightarrow \nu_{\mu})$
as a function of $L_{\nu}/E_{\nu}$
under the neutrino oscillation parameters
obtained by Super-Kamiokande Collaboration
.}
\label{figJ017}
\end{center}
\end{figure}

\section{$L/E$ Distributions in Our Numerical Computer Experiment}

Our computer numerical experiments are carried out in the unit of
1489.2 days. Hereafter, we call 1489.2 live days 
 as one SK live day. 
The live days of 1489.2  is the total live days for the analysis of the 
neutrino events generated inside the detector by Super-Kamiokande 
Collaboration \cite{Ashie2}.
We repeat one SK live day experiment as much as 25 times,
namely, the total live days for our computer numerical experiments 
is 37230 live days (25 SK live days).
 In Figure~\ref{figJ015}, we show $L_{\nu}/E_{\nu}$ 
distribution without oscillation for one experiment (1489.2 live days)
 among twenty five computer numerical experiments.
In those numerical experiments, there are statistical uncertainties only
which are due to both the stochastic character in the physical
 processes concerned and the geometry of the detectors.
Therefore we add the standard deviation as for the statistical 
uncertainty around their average in the forthcoming graphs, if neccessary. 
 In Figure~\ref{figJ016}, we show the 
statistical uncertainty, the standard deviations around their average 
values through twenty five experiments.    
Similarly for
other possible combinations of $L$ and $E$ ( 
$L_{\nu}/E_{\mu}$, $L_{\mu}/E_{\nu}$ and $L_{\mu}/E_{\mu}$) for 
37230 live days (25 SK live days) we did so.
  
\subsection{$L_{\nu}/E_{\nu}$ distribution}
\subsubsection{For null oscillation}

In Figures~\ref{figJ015} and ~\ref{figJ016},
both distributions show the sinusoidal-like
 character for $L_{\nu}/E_{\nu}$ distribution, namely, 
the appearance of the top and the bottom, even for null
 oscillation. The uneven histograms in Figure~\ref{figJ015}, comparing 
 with those in Figure~\ref{figJ016}, show that the statistics of 
 Figure~\ref{figJ015} is not enough compared with that of 
Figure~\ref{figJ016}.
Roughly speaking, smaller $L_{\nu}/E_{\nu}$ 
correspond to the contribution from downward neutrinos, 
larger $L_{\nu}/E_{\nu}$ correspond to that from upward neutrinos 
and $L_{\nu}/E_{\nu}$ near the minimum 
correspond to the horizontal neutrinos, although the real situation is
more complicated, because the backscattering effect in QEL
as well as the azimuthal angle effect in QEL
could not be neglected.
From Figure~\ref{figJ016}, we understand that the bottom around
 70~km/GeV
denotes the contribution from the horizontal direction
and has no relation with neutrino oscillation in any sence.

\begin{figure}
\begin{center}
\resizebox{0.45\textwidth}{!}{%
  \includegraphics{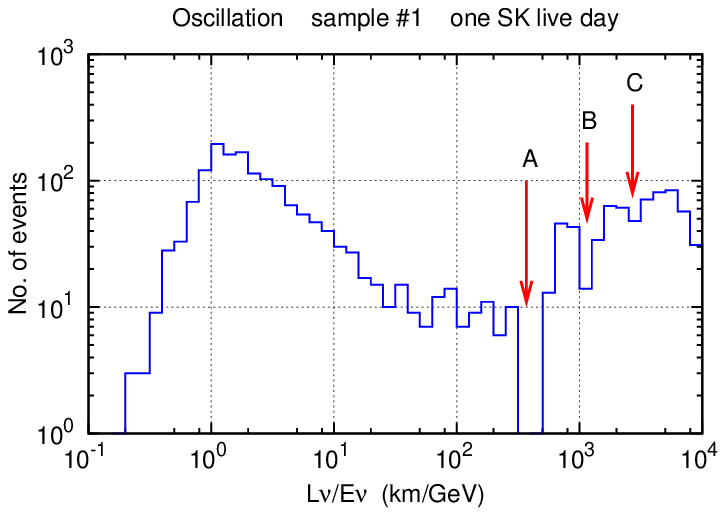}
}
\caption{$L_{\nu}/E_{\nu}$ distribution with oscillation
for 1489.2 live days (one SK live day), sample No.1.}
\label{figJ018}
\resizebox{0.45\textwidth}{!}{%
  \includegraphics{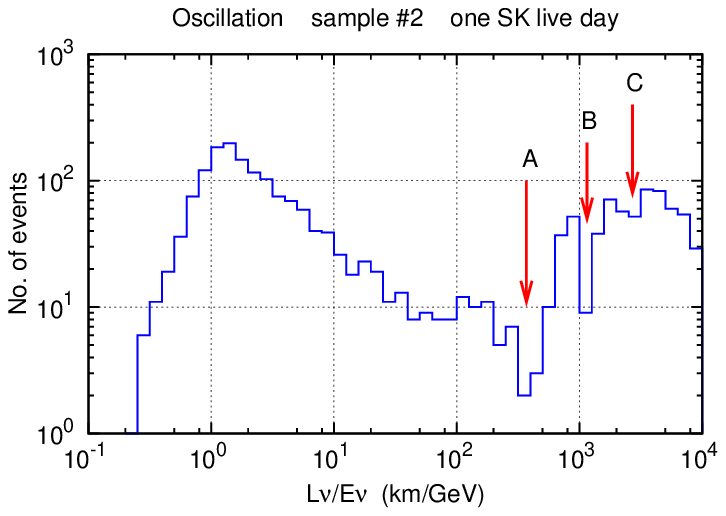}
}
\caption{$L_{\nu}/E_{\nu}$ distribution with oscillation
for 1489.2 live days (one SK live day), sample No.2.}
\label{figJ019}      
\resizebox{0.45\textwidth}{!}{%
  \includegraphics{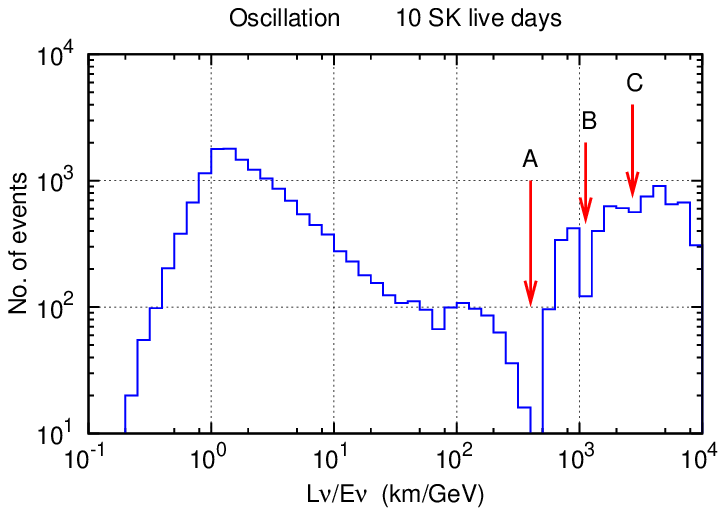}
}
\caption{$L_{\nu}/E_{\nu}$ distribution with oscillation
for 14892 live days (10 SK live days).}
\label{figJ020}       
\end{center}
\end{figure}

\begin{figure}
\begin{center}
\resizebox{0.45\textwidth}{!}{%
  \includegraphics{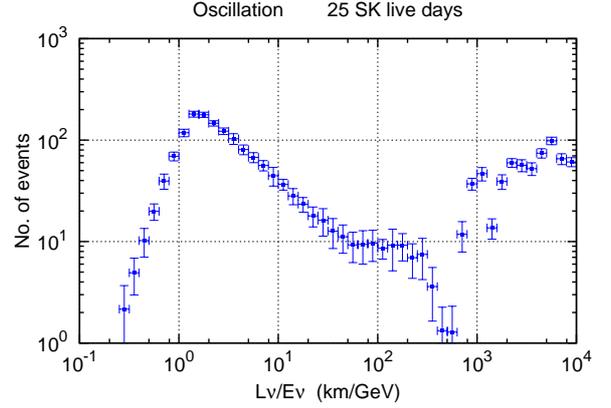}
}
\caption{$L_{\nu}/E_{\nu}$ distribution with standard deviations
with oscillation for 37230 live days (25 SK live days).}
\label{figJ021}
\end{center}
\end{figure}

\begin{figure}
\begin{center}
\resizebox{0.5\textwidth}{!}{%
  \includegraphics{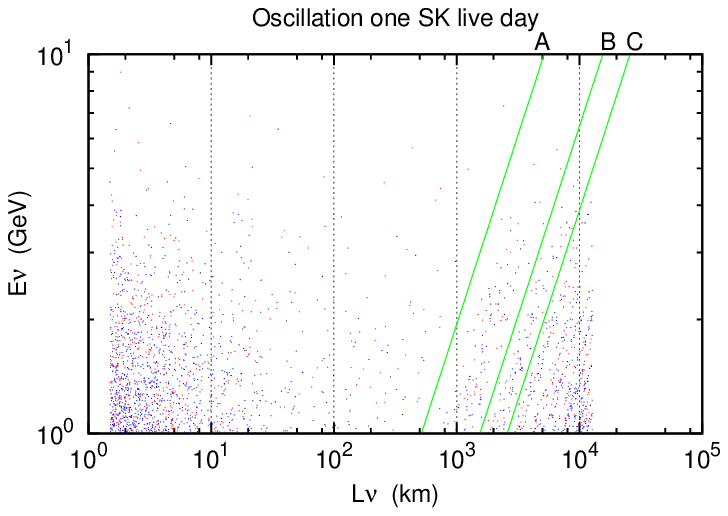}
}
\caption{The correlation diagram between $L_{\nu}$ and $E_{\nu}$
with oscillation for 1489.2 live days (one SK live day).}
\label{figJ022}     
\resizebox{0.5\textwidth}{!}{%
  \includegraphics{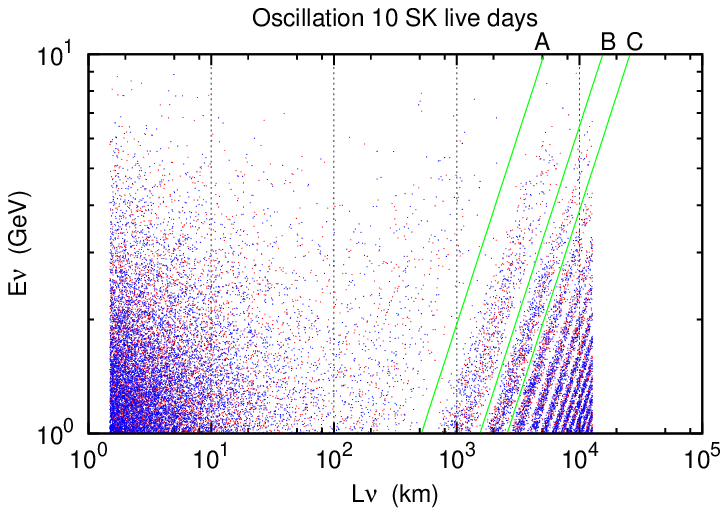}
}
\caption{The correlation diagram between $L_{\nu}$ and $E_{\nu}$
with oscillation for 14892 live days (10 SK live days).}
\label{figJ023}       
\vspace{-1cm}
\hspace*{-1.0cm}
\rotatebox{90}{%
\resizebox{0.4\textwidth}{!}{%
  \includegraphics{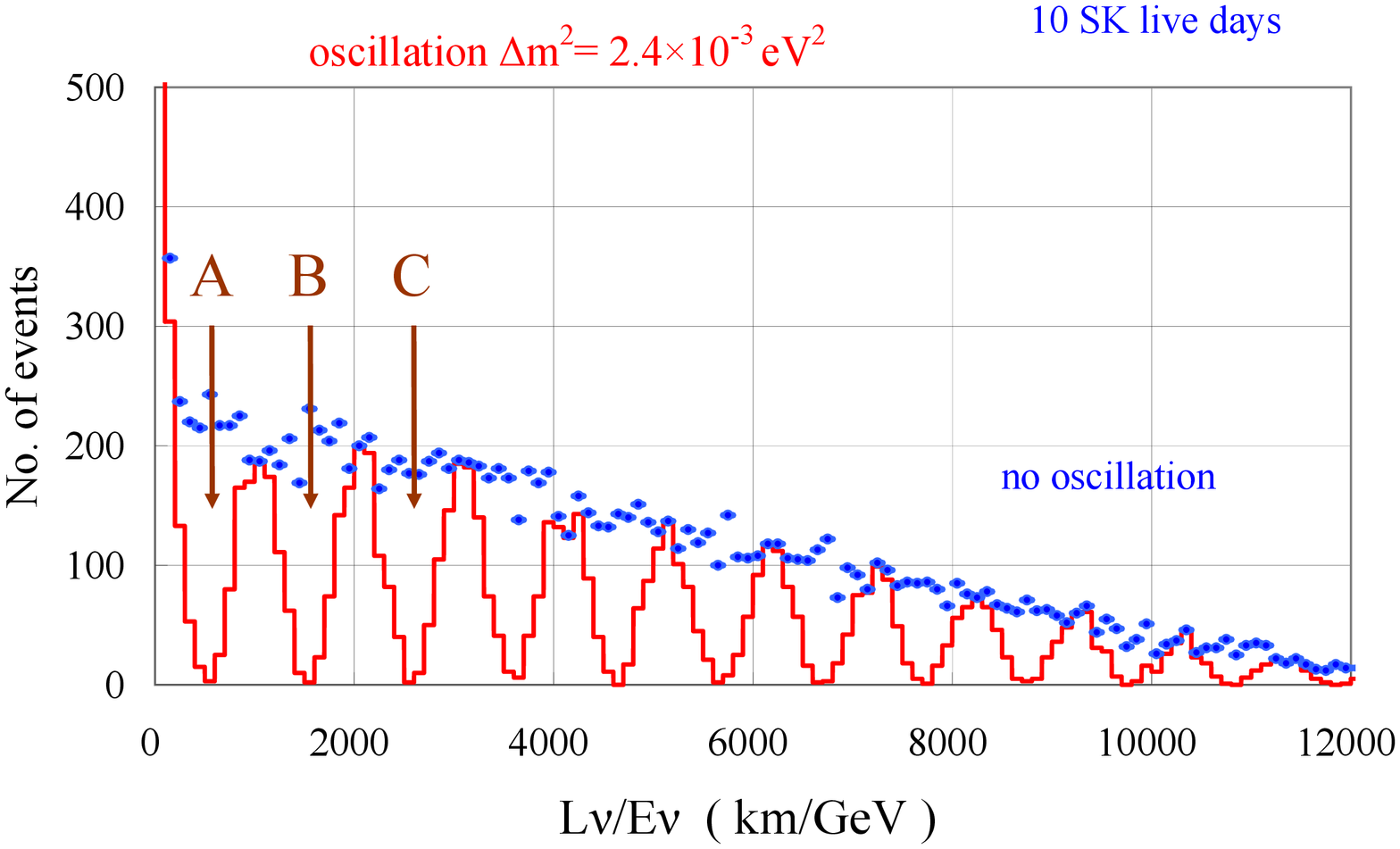}
}}
\vspace{-1.5cm}
\caption{$L_{\nu}/E_{\nu}$ distribution
with and without oscillation for 14892 live days (10 SK live days).}
\label{figJ024}       
\end{center}
\end{figure}

\begin{figure}
\begin{center}
\resizebox{0.45\textwidth}{!}{%
  \includegraphics{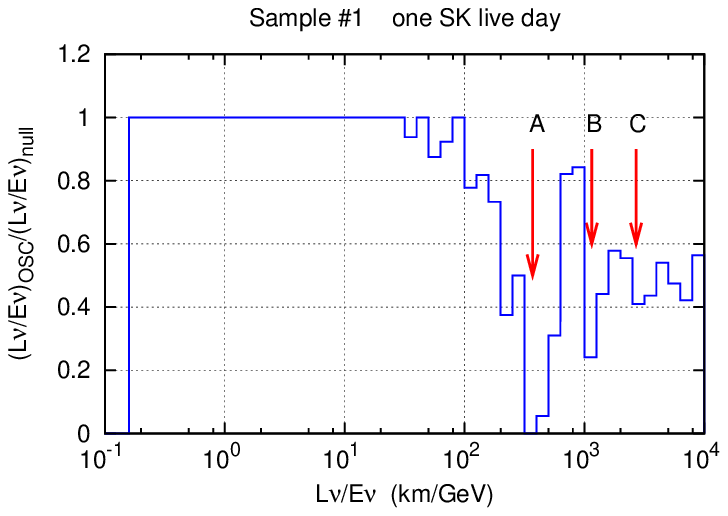}
}
\caption{The ratio of 
$(L_{\nu}/E_{\nu})_{osc}/(L_{\nu}/E_{\nu})_{null}$
 for 1489.2 live days (one SK live day).}
\label{figJ025}       

\resizebox{0.45\textwidth}{!}{%
  \includegraphics{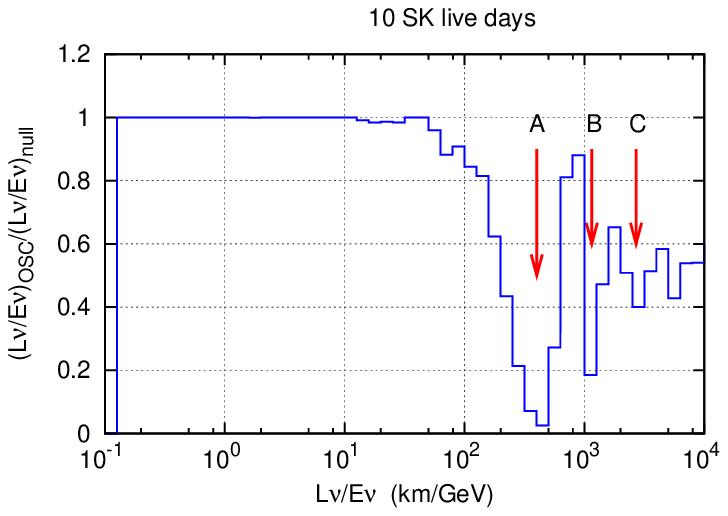}
}
\caption{The ratio of 
$(L_{\nu}/E_{\nu})_{osc}/(L_{\nu}/E_{\nu})_{null}$
 for 14892 live days (10 SK live days).}
\label{figJ026}       
\end{center}
\end{figure}

\subsubsection{For oscillation (SK oscillation parameters)}

The survival probability of a given flavor, such as $\nu_{\mu}$,
 is given by
$$
P(\nu_{\mu} \rightarrow \nu_{\mu}) 
= 1- sin^2 2\theta \cdot sin^2
(1.27\Delta m^2 L_{\nu} / E_{\nu} ).         \,\,\,(1)  
$$                                                    

\noindent Then, for maximum oscillations under SK neutrino oscillation 
parameters,  we have
$$
 1.27\Delta m^2 L_{\nu} / E_{\nu} = (2n+1)\times\frac{\pi}{2},  \,\,\,\,\,\,(2)  
$$                                                    
where $\Delta m^2 = 2.4\times 10^{-3}\rm{eV^2}$.
From Eq.(2), we have the following values of
$L_{\nu} / E_{\nu}$ for maximum oscillations.
\begin{eqnarray}
     L_{\nu}/E_{\nu}
  &=& 515 {\rm km/GeV}\,\,\,\,\, for\,\, n=0   \,\,\, (3-1)\nonumber\\
 &=& 1540 {\rm km/GeV}\,\,\,   for\,\, n=1   \,\,\, (3-2)\nonumber\\ 
 &=& 2575 {\rm km/GeV}\,\,\,   for\,\, n=2   \,\,\, (3-3) \nonumber\\
&{\rm and}\,{\rm so}\,{\rm on.}\nonumber\\
\nonumber
\end{eqnarray}
In Figure~\ref{figJ017}, we give the survival probability 
$P(\nu_{\mu} \rightarrow \nu_{\mu})$ as a function of 
$L_{\nu} / E_{\nu}$ under the neutrino oscillation parameters obtained 
by Super-Kamiokande Collaboration.
 In cosmic ray experiments, the energy spectrum of the incident neutrino, 
is convoluted into the survival probability.

In Figure~\ref{figJ018}, we give one example of the $L_{\nu} / E_{\nu}$ 
distribution for one SK live day (1489.2 live days)\cite{Ashie2}
among twenty five sets of 
the computer numerical experiments in the unit of one SK live day. 
In Figure~\ref{figJ019}, we give another example for one SK 
live day. Arrows A, B and C represent 
the first, the second and the third maximum oscillation which are 
given in Eq. (3-1), (3-2) and (3-3), respectively. By the definition 
of our computer numerical experiments, there are 
no experimental error bars 
in $L_{\nu} / E_{\nu}$ distributions in Figures~\ref{figJ018} and 
\ref{figJ019}.

 In Figure~\ref{figJ020}, we show the $L_{\nu} / E_{\nu}$
 distribution for 14892 live 
days (10 SK live days). Compared 
Figure~\ref{figJ020}20 with Figures~\ref{figJ018} and \ref{figJ019},
  it is clear that  
$L_{\nu} / E_{\nu}$ distribution in Figure~\ref{figJ020} becomes 
smoother due to larger statistics. 
In Figure~\ref{figJ021}, 
we can add the statistical uncertainty (standard deviation 
in this case) around their average, because every one SK live 
day experiment among twenty five sets of the experiments 
fluctuates one by one due to their stochastic character 
in their physical processes and geometrical conditions 
of the detectors concerned. 
 In order to make the image of the maximum 
oscillations in $L_{\nu} / E_{\nu}$
 distributions clearer, we show the 
correlations between $L_{\nu}$ and $E_{\nu}$
 in Figures~\ref{figJ022} and \ref{figJ023}, 
which correspond to Figures~\ref{figJ018} and \ref{figJ020},
 respectively.
 In Figure~\ref{figJ022} for one SK live day,
we can observe vacant regions for  the events concerned 
assigned by A, B and C. 
In Figure~\ref{figJ023} for ten SK live days, the existence of 
the vacant regions for the events concerned becomes 
 clearer due to larger statistics. 

In Figure~\ref{figJ024}, we give $L_{\nu} / E_{\nu}$ distribtution
 with 14892 live 
days (10 SK live days) in the linear scale which is another 
expression of the same content as in Figure~\ref{figJ023}.
 Also, it is the survival probability convoluted with the incident 
neutrino energy spectrum. If we compare Figure~\ref{figJ024} 
with Figure~\ref{figJ017},
 then, we clearly see the series of maximum oscillations 
characterized with n=0 (A), 1(B), 2(C) and so on which are given by 
Eq.(2). It is clear from Figure~\ref{figJ024} that the maximum 
oscillations with 
n=0,1 and 2 have the almost same frequencies
\footnote{
Super-Kamiokande Collaboration never mentioned 
existence of the second and the third maximum oscillations (n=1 and 2)}
 under the incident neutrino 
energy spectrum utilized by Super-Kamiokande Collaboration \cite{Honda}
(see footnote 1).  
The situation shown in Figures~\ref{figJ022}
 to \ref{figJ024} shows definitely that our computer 
numerical experiment are carried out as exactly as possible from the view 
point of the stochastic treatment to the matter. 

 We have repeated the 
computer numerical experiment for one SK live day as 
much as twenty five times independently,
in both cases with oscillation and without 
oscillation.  Consequently, there are 625 ($=25\times25$) 
sets of ratios of
$(L_{\nu}/E_{\nu})_{osc}/(L_{\nu}/E_{\nu})_{null}$ 
for one SK live day which correspond 
to Eq.(1). In Figure~\ref{figJ025}, we show one example among 625 
combinations.
 In Figure~\ref{figJ026}, we show the same ratio 
for 14892 live days (10 SK live days).
 In conclusion, from Figures~\ref{figJ018} to \ref{figJ026},
 we can reproduce the minimum extrema for neutrino oscillation in 
our $L_{\nu} / E_{\nu}$ analysis.
This fact shows doubtlessly 
that our computer numerical experiments are done
in the correct manner.

\begin{figure}
\begin{center}
\resizebox{0.45\textwidth}{!}{%
  \includegraphics{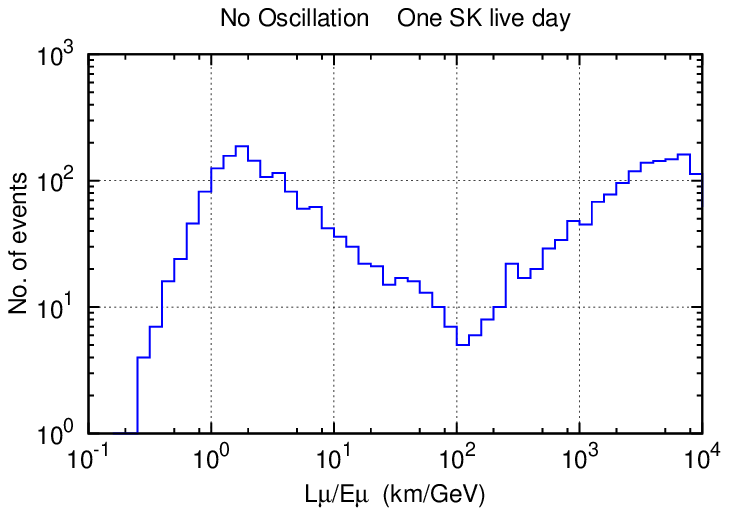}
}
\caption{$L_{\mu}/E_{\mu}$ distribution without oscillation
for 1489.2 live days (one SK live day).}
\label{figJ027}       
\resizebox{0.45\textwidth}{!}{%
  \includegraphics{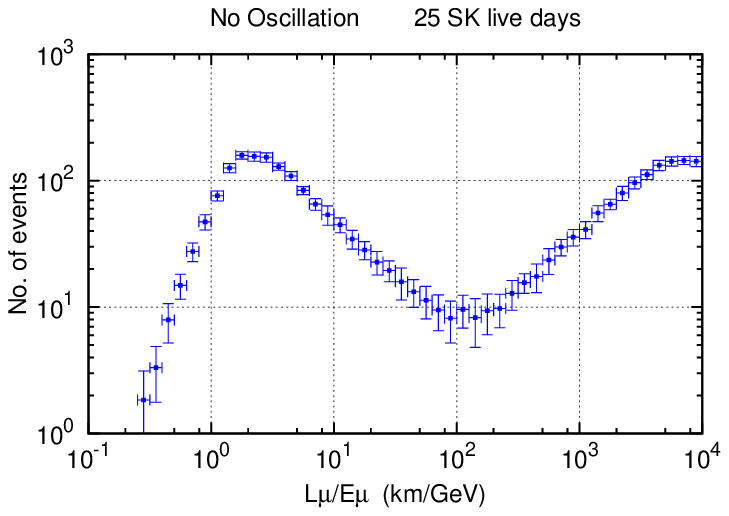}
}
\caption{$L_{\mu}/E_{\mu}$ distribution without oscillation
for 37230 live days (25 SK live days).}
\label{figJ028}
\end{center}
\end{figure}

\begin{figure}
\begin{center}
\resizebox{0.45\textwidth}{!}{%
  \includegraphics{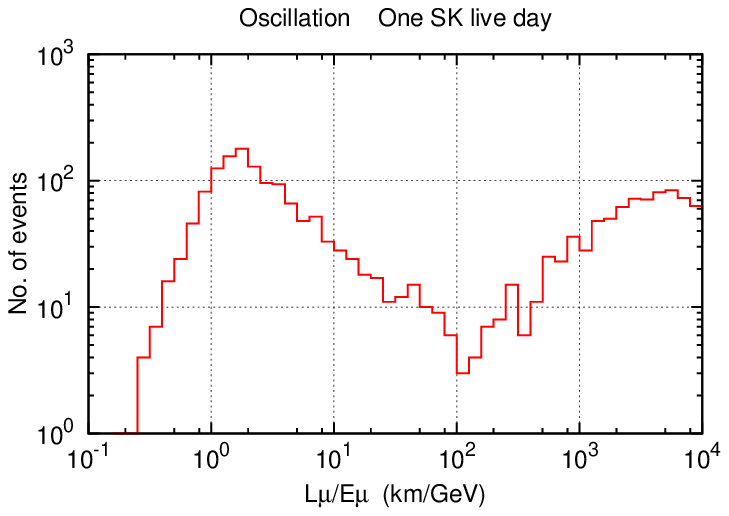}
}
\caption{$L_{\mu}/E_{\mu}$ distribution with oscillation
for 1489.2 live days (one SK live day).}
\label{figJ029}       
\resizebox{0.45\textwidth}{!}{%
  \includegraphics{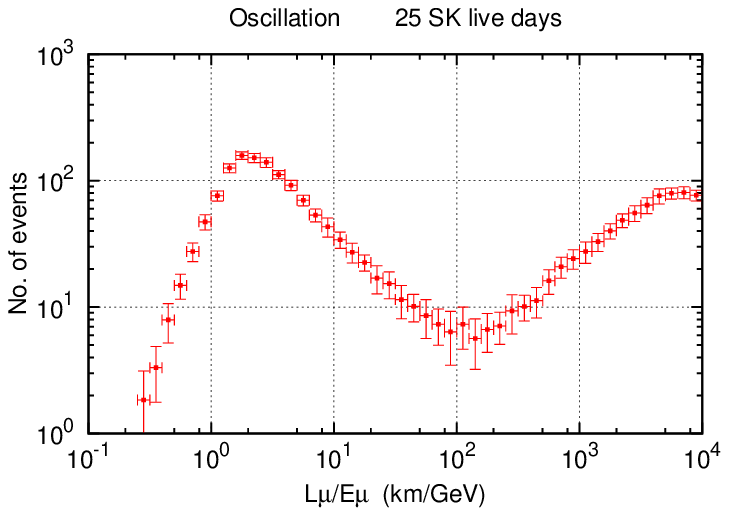}
}
\caption{$L_{\mu}/E_{\mu}$ distribution with oscillation
for 37230 live days (25 SK live days).}
\label{figJ030}
\end{center}
\end{figure}

\subsection{$L_{\mu}/E_{\mu}$ distribution}
As physical quantities which can really be observed are
$L_{\mu}$ and $E_{\mu}$ instead of $L_{\nu}$ and $E_{\nu}$,
therefore we examine $L_{\mu}/E_{\mu}$ distribution
focusing the existence of the maximum oscillation.

\subsubsection{For null oscillation}
 In Figure~\ref{figJ027}, we give one sample for one SK live day
 (1489.2 live days) from the totally 37230 live days (25 SK live days) 
events, each of which has 1489.2 live days.
 Figure~\ref{figJ028} shows the average distribution accompanied by the 
statistical uncertainty bar (not experimental error bar).
 It is clear from these figures that the existence of the dip or bottom, 
namely
the sinusoidal character, means the contribution merely from horizontal 
contribution, having no relation with any neutrino oscillation 
character.   

\begin{figure}
\begin{center}
\resizebox{0.45\textwidth}{!}{%
  \includegraphics{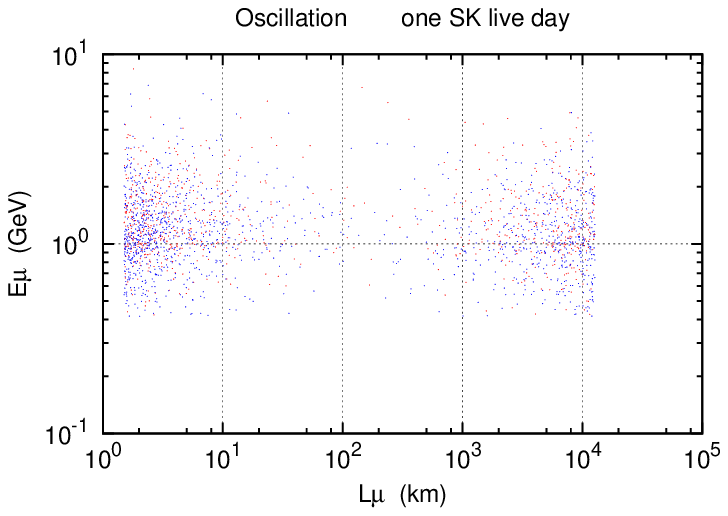}
}
\caption{The correlation diagram between $L_{\mu}$ and $E_{\mu}$
with oscillation for 1489.2 live days (one SK live day).}
\label{figJ031}
\resizebox{0.45\textwidth}{!}{%
  \includegraphics{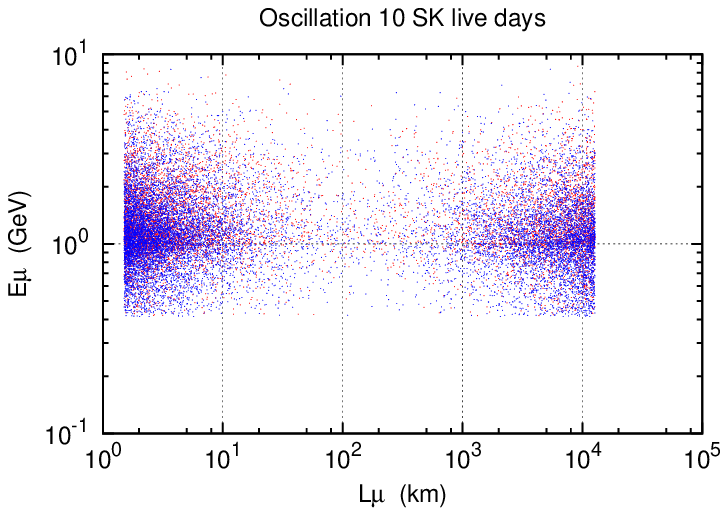}
}
\caption{The correlation diagram between $L_{\mu}$ and $E_{\mu}$
with oscillation for 14892 live days (10 SK live days).}
\label{figJ032}
\vspace{-1cm}
\hspace*{-1cm}
\rotatebox{90}{%
\resizebox{0.4\textwidth}{!}{%
  \includegraphics{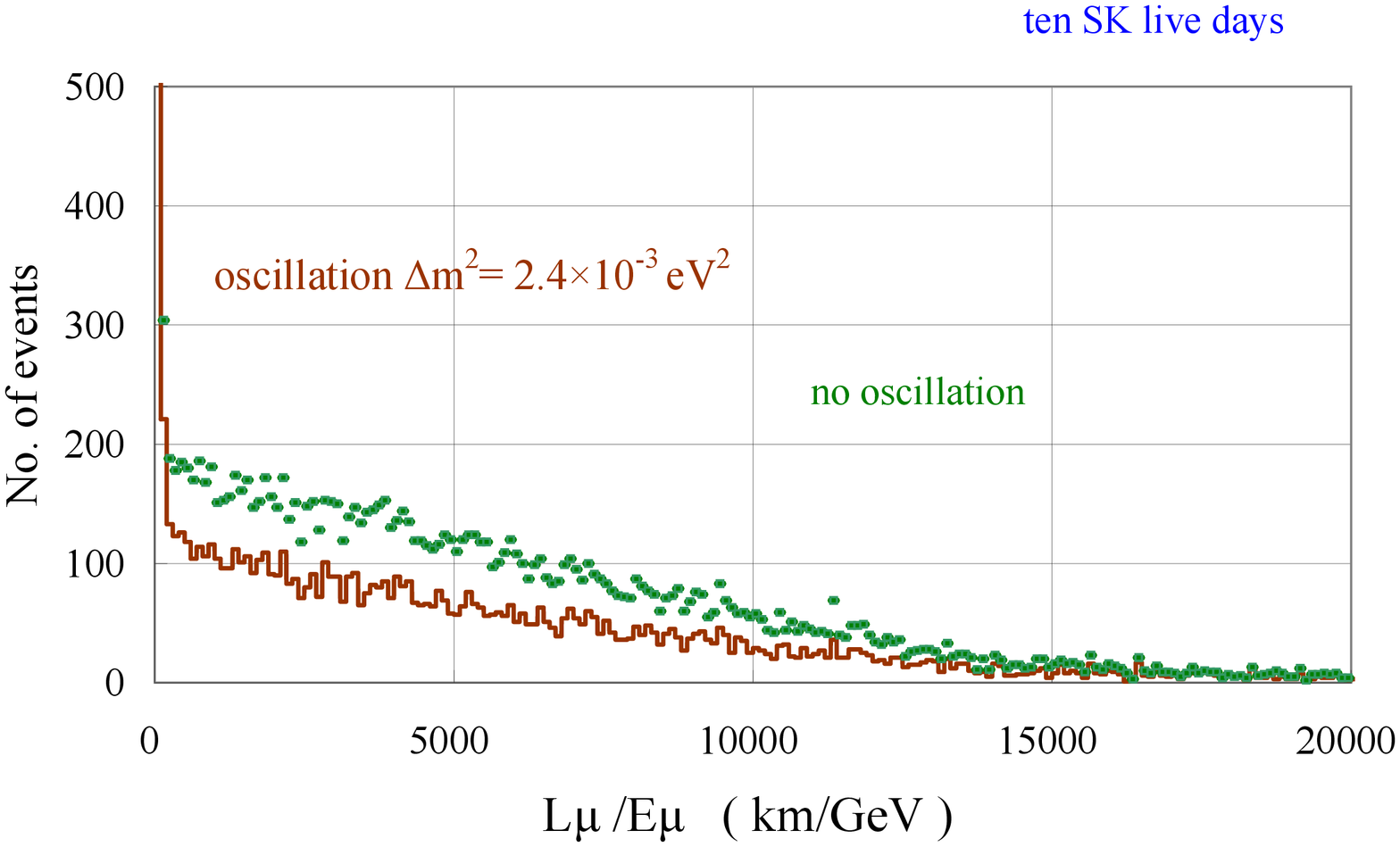}
}}
\vspace{-1cm}
\caption{$L_{\mu}/E_{\mu}$ distribution
with and without oscillation for 14892 live days (10 SK live days).}
\label{figJ033}

\end{center}
\end{figure}

\begin{figure}
\begin{center}
\resizebox{0.45\textwidth}{!}{%
  \includegraphics{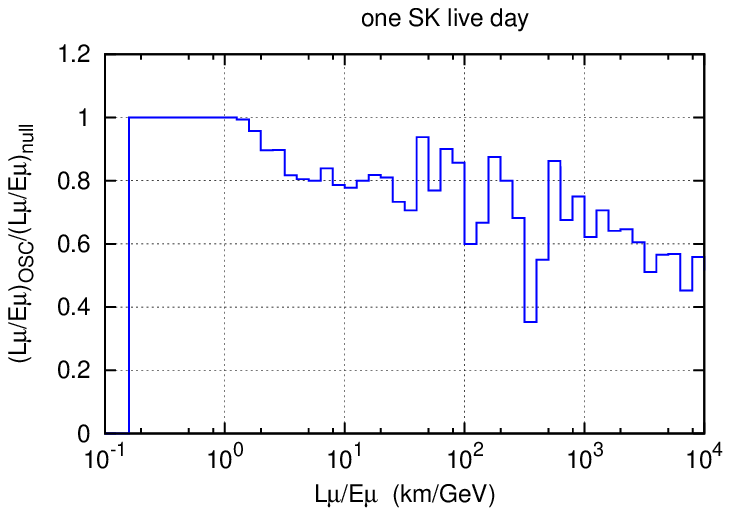}
}
\caption{The ratio of 
$(L_{\mu}/E_{\mu})_{osc}/(L_{\mu}/E_{\mu})_{null}$
 for 1489.2 live days (one SK live day).}
\label{figJ034}

\resizebox{0.45\textwidth}{!}{%
  \includegraphics{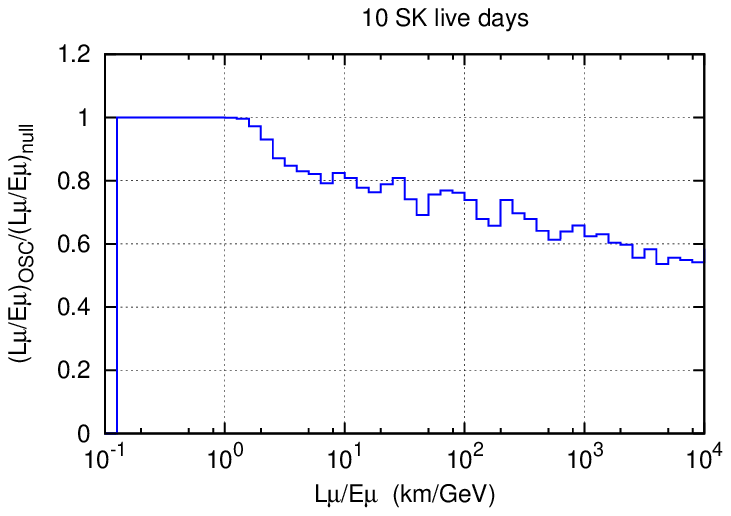}
}
\caption{The ratio of 
$(L_{\mu}/E_{\mu})_{osc}/(L_{\mu}/E_{\mu})_{null}$
 for 14892 live days (10 SK live days).}
\label{figJ035}
\end{center}
\end{figure}

\begin{figure}
\begin{center}
\resizebox{0.45\textwidth}{!}{%
  \includegraphics{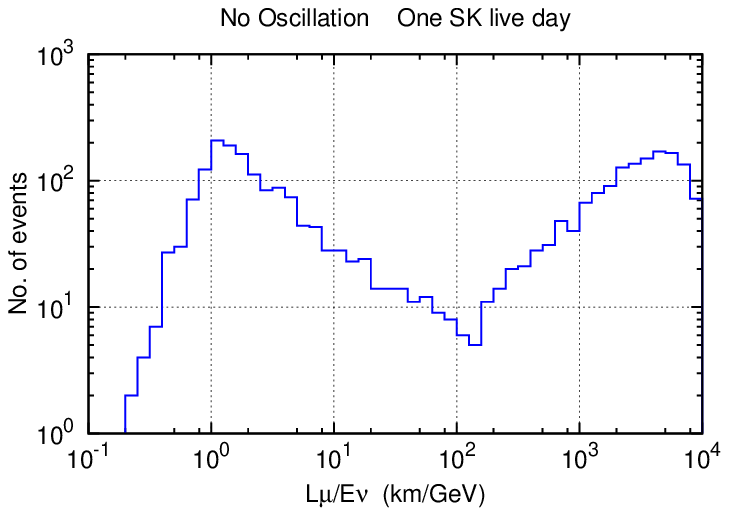}
}
\caption{$L_{\mu}/E_{\nu}$ distribution without oscillation
for 1489.2 live days (one SK live day).}
\label{figJ036}
\resizebox{0.45\textwidth}{!}{%
  \includegraphics{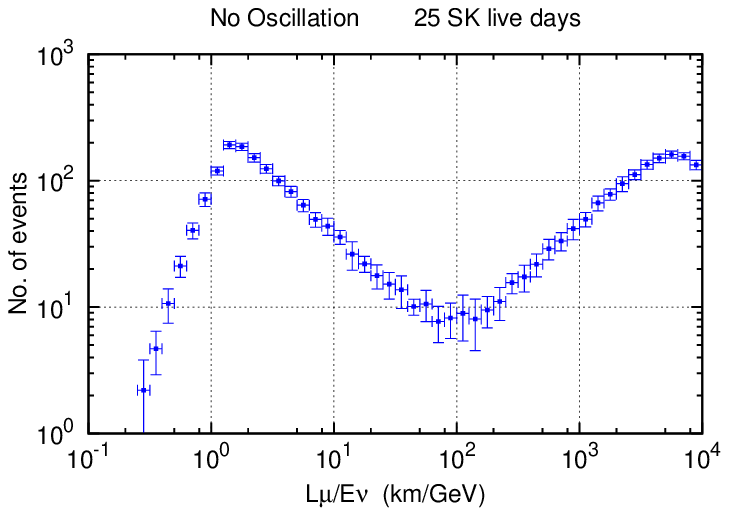}
}
\caption{$L_{\mu}/E_{\nu}$ distribution without oscillation
for 37230 live days (25 SK live days).}
\label{figJ037}
\end{center}
\end{figure}
\subsubsection{For oscillation (SK oscillation parameters)}
In Figures~\ref{figJ029} and \ref{figJ030}, 
we give the $L_{\mu}/E_{\mu}$ distributions with 
 oscillation for 1489.2 live days (one SK live day) and 37230 live days 
(25 SK live days), respectively. 
In Figure~\ref{figJ029}, we may observe the uneven histogram,
 something like
 curious bottoms coming from neutrino oscillation. However, 
in Figure~\ref{figJ030} where 
the statistics is 25 times as much as that of 
Figure~\ref{figJ029}, the histogram
 becomes smoother and such bottoms disappear, which 
turns out finally for the bottoms to be pseudo. 
It is impossible to extract the neutrino oscillation parameters from the 
comparison of Figure~\ref{figJ030} with Figure~\ref{figJ028}.

In Figures~\ref{figJ031} and \ref{figJ032}, correspondingly,
 we give the correlation 
between $L_{\mu}$ and $E_{\mu}$ for 1489.2 live days (one SK live day)
and 14892 live days (10 SK live days), respectively.

In Figure~\ref{figJ033}, we give the $L_{\mu}/E_{\mu}$ distribution 
for 14892 live days (10 SK live days) in the linear scale which is 
another expression of the same content as in Figure~\ref{figJ032}. 
As in Figure~\ref{figJ032}, we cannot find any maximum oscillation-like 
phenomena in Figure~\ref{figJ033}, which is contrast to 
Figure~\ref{figJ024}.

 It is clear from the figures that we can not
observe the maximum oscillation in $L_{\mu}/E_{\mu}$, 
on the contrary to Figures~\ref{figJ018} to \ref{figJ024}
which give the maximum oscillations.
 Namely, we may conclude that we can not observe the 
sinusoidal flavor transition probability of neutrino oscillation
against the claim by Super-Kamiokande Collaboration\cite{Ashie1}
when we adopt physically observable quantities, such as
 $L_{\mu}$ and $E_{\mu}$.\\
\begin{figure}
\begin{center}
\resizebox{0.45\textwidth}{!}{%
  \includegraphics{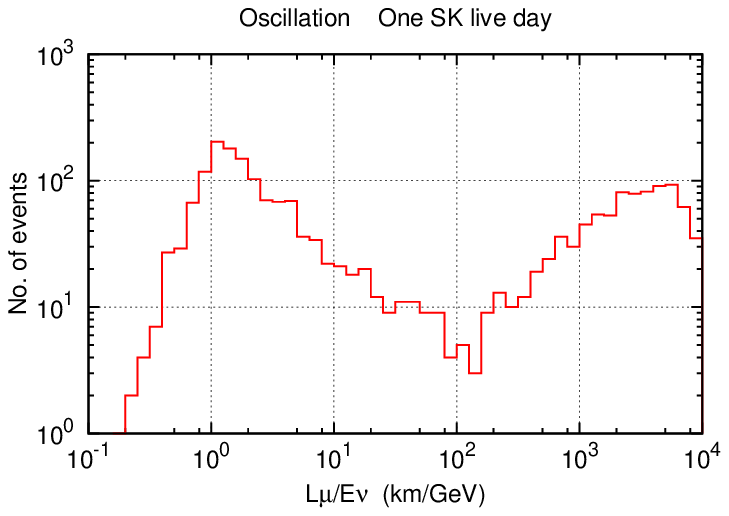}
}
\caption{$L_{\mu}/E_{\nu}$ distribution with oscillation
for 1489.2 live days (one SK live day).}
\label{figJ038}
\resizebox{0.45\textwidth}{!}{%
  \includegraphics{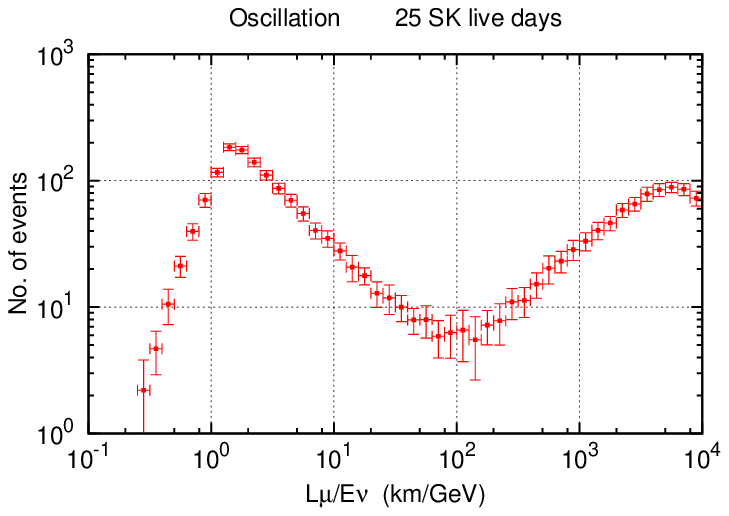}
}
\caption{$L_{\mu}/E_{\nu}$ distribution with oscillation
for 37230 live days (25 SK live days).}
\label{figJ039}
\end{center}
\end{figure}
\begin{figure}
\begin{center}
\resizebox{0.45\textwidth}{!}{%
  \includegraphics{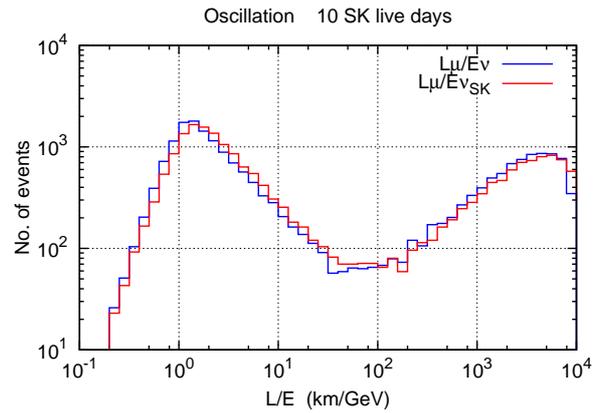}
}
\caption{The $L_{\mu}/E_{\nu,SK}$ distribution in
comparison with $L_{\mu}/E_{\nu}$ distribution with oscillation 
for 14892 day (10 SK live days).}
\label{figJ040}
\end{center}
\end{figure}
\begin{figure}
\begin{center}
\resizebox{0.45\textwidth}{!}{%
  \includegraphics{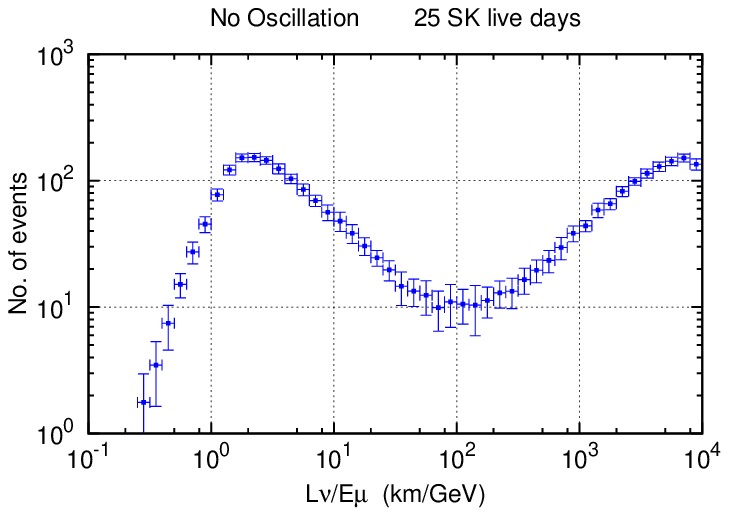}
}
\caption{The $L_{\nu}/E_{\mu}$ distribution without
oscillation for 37230 days (25 SK live days).}
\label{figJ041}
\resizebox{0.45\textwidth}{!}{%
  \includegraphics{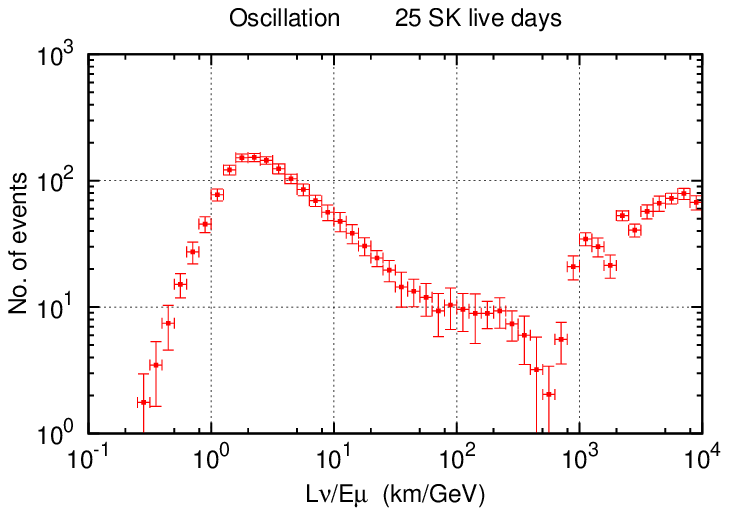}
}
\caption{The $L_{\nu}/E_{\mu}$ distribution with 
oscillation for 37230 days (25 SK live days).}
\label{figJ042}
\vspace{-1cm}
\end{center}
\end{figure}

In order to confirm the disappearance of the psuedo maximum 
oscillations, in Figures~\ref{figJ034} and \ref{figJ035}, we give
the survival probability of a given flavor 
for $L_{\mu}/E_{\mu}$ distribution, namely,
$(L_{\mu}/E_{\mu})_{osc}/(L_{\mu}/E_{\mu})_{null}$,
 for 1489.2 live days (one SK live day)
and 14892 live days (10 SK live days), respectively.
In Figure~\ref{figJ034}, we show one example of
$(L_{\mu}/E_{\mu})_{osc}/(L_{\mu}/E_{\mu})_{null}$
among 625 sets of ratios.
 Comparing Figure~\ref{figJ034} with Figure~\ref{figJ035},
the pseudo dips in Figure~\ref{figJ034} disappear in Figure~\ref{figJ035}.
Thus the histogram becomes a rather decreasing function of 
$L_{\mu}/E_{\mu}$ in Figure~\ref{figJ035}. 
If we further make statistics higher, the survival probability for
$L_{\mu}/E_{\mu}$ distribution should be a monotonously decreasing 
function of $L_{\mu}/E_{\mu}$, whithout showing any 
characteristics of the maximum oscillation,  
which is in contrast to Figures~\ref{figJ025} and \ref{figJ026}.

 In conclusion, we should say that we can not find any maximum 
oscillation for the neutrino oscillation in the 
$L_{\mu}/E_{\mu}$ distribution.

\subsection{$L_{\mu}/E_{\nu}$ distribution}
Now, we examine the $L_{\mu}/E_{\nu}$ distribution which 
 Super-Kamiokande Collaboration treat 
in the thier 
paper, expecting the evidence for the oscillatory signatuture in 
atmospheric neutrino oscillations.
\subsubsection{For null oscillation}
In Figures~\ref{figJ036} and \ref{figJ037},
 we give the $L_{\mu}/E_{\nu}$ distribution without 
oscillation for 1489.2 live days (one SK live day) and 37230 live days
(25 SK live days), respectively.
Comparing Figure~\ref{figJ036} with Figure~\ref{figJ037},
 the larger statistics makes the
 distribution smoother.
 Also, there is a sinusoidal-like bottom which has no relation with 
neutrino oscillation.
\subsubsection{For oscillation (SK oscillation parameters)}
In Figures~\ref{figJ038} and \ref{figJ039}, 
we give the $L_{\mu}/E_{\nu}$ distribution with 
oscillation for 1489.2 live days (one SK live day) and 37230 live days
(25 SK live days), respectively. 
In Figure~\ref{figJ038},
 we may find something like a bottom which corresponds to the 
first maximum oscillation near $\sim$200 (km/GeV).
However, such the dip disappears, by making the statistics larger as 
shown in Figure~\ref{figJ039}.

\subsubsection{$L_{\mu}/E_{\nu,SK}$ distribution for the oscillation}
 Instead of $E_{\nu}$ which is correctly sampled from the corresponding 
probability functions, 
let us utilize $E_{\nu,SK}$ which is obtained from the "approximate" 
formula (Eq.(6) in the preceding paper\cite{Konishi2}).

 We express $E_{\nu}$ described in Eq.(6) of the preceding 
paper\cite{Konishi2}
utilized by Super-Kamiokande Collaboration as $E_{\nu,SK}$ 
to discriminate our $E_{\nu}$ obtained in the stochastic manner 
correctly.

In Figure~\ref{figJ040}, 
we give $L_{\mu}/E_{\nu,SK}$ distribution for 14892 live days 
(10 SK live days),
comparing with $L_{\mu}/E_{\nu}$ distribution.
It is understood from the comparison that there is no significant 
difference between $L_{\mu}/E_{\nu,SK}$ distribution and 
$L_{\mu}/E_{\nu}$ one.
This fact tells us that the "aproximate" formula for $E_{\nu}$ by 
Super-Kamiokande Collaboration 
does not produce so significant error practically.
Although this kind of foumula is not suitable for the treatment of
stochastic quantities,
the result is understandable from Figure~14 in the preceding
 paper\cite{Konishi2}.
 Also, we can conclude that we do not 
find any hole corresponding to the maximum oscillation in 
$L_{\mu}/E_{\nu}$ or $L_{\mu}/E_{\nu,SK}$ distributions.
The reason why the Figure~\ref{figJ039} can not show such dip structure 
 as shown in Figures~\ref{figJ018} and \ref{figJ019},  
 comes from the situation that the role of $L_{\nu}$
is much more crucial than that of $E_{\nu}$ in the $L/E$ analysis.
Namely, $L_{\nu}$ cannot be replaced by $L_{\mu}$ at all.
 Also, see the discussion in the following subsection 4.4.    
    
\begin{figure}
\begin{center}
\resizebox{0.45\textwidth}{!}{%
  \includegraphics{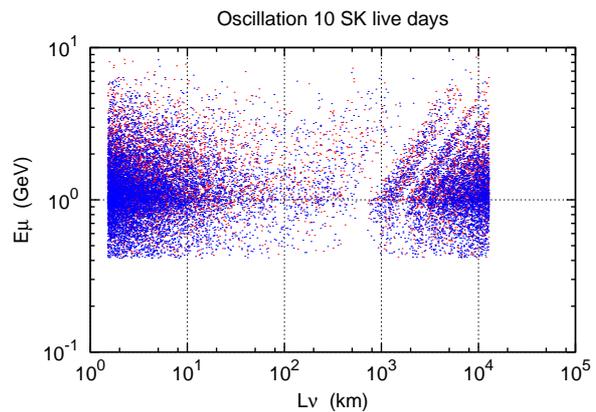}
}
\caption{The correlation diagram between $L_{\nu}$ and $E_{\mu}$
 with oscillation for 14892 days (10 SK live days).}
\label{figJ043}
\end{center}
\end{figure}

\subsection{$L_{\nu}/E_{\mu}$ distribution}
\subsubsection{For null oscillation}
 In Figure~\ref{figJ041}, we give $L_{\nu}/E_{\mu}$ distribution without 
oscillation for 37230 live days (25 SK live days)
of Super-Kamiokande Experiment
 to consider statistical fluctuation effect as precisely as possible. 
It is clear from the figure that there is not any dip 
corresponding to 
the maximum oscillation which is expected to appear in presence of 
 neutrino oscillation, as it must be. 
\subsubsection{For oscillation (SK oscillation parameters)}
In Figure~\ref{figJ042}, we give the corresponding  
distribution with the oscillation. In Figure~\ref{figJ043},
 we give the correlation 
diagram between $L_{\nu}$ and $E_{\mu}$
for 14823 live days (10 SK live days). 
On the contrary to Figure~\ref{figJ039}, there are surely 
some kinds of holes in Figure~\ref{figJ042},
 and furthermore we can discriminate the strip pattern in 
Figure~\ref{figJ043}, similarly as in Figure~\ref{figJ023}.

\begin{figure}
\begin{center}
\resizebox{0.45\textwidth}{!}{%
  \includegraphics{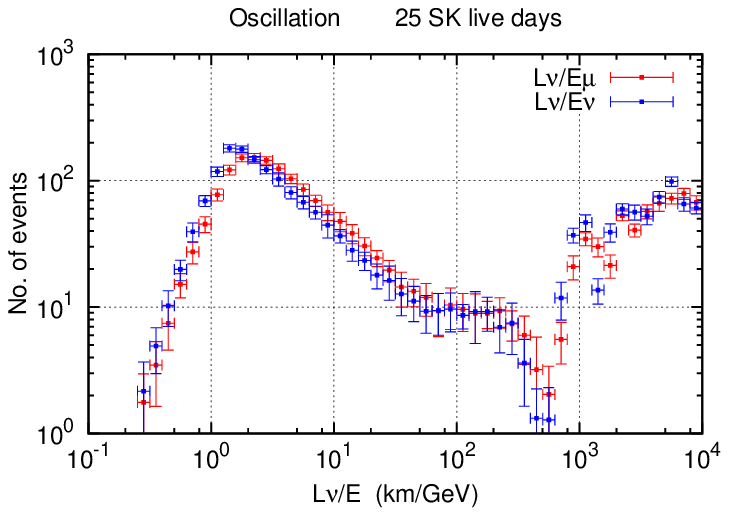}
}
\caption{Comparison between 
$L_{\nu}/E_{\nu}$ distribution and $L_{\nu}/E_{\mu}$ distribution 
 with oscillation for 37230 days (25 SK live days).}
\label{figK044}
\resizebox{0.45\textwidth}{!}{%
  \includegraphics{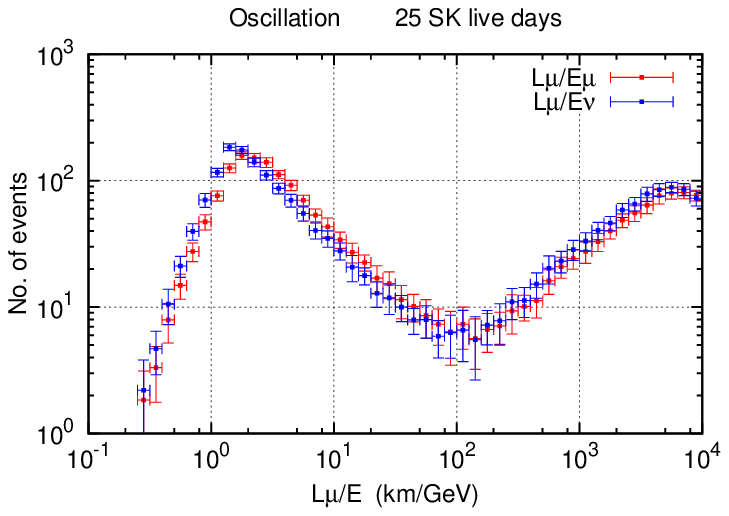}
}
\caption{Comparison between 
$L_{\mu}/E_{\nu}$ distribution and $L_{\mu}/E_{\mu}$ distribution 
 with oscillation for 37230 days (25 SK live days).}
\label{figK045}
\end{center}
\end{figure}

Therefore, we surmise from Figures~\ref{figJ042} and 
\ref{figJ043} that we may observe 
some "maximum oscillation like" quantities
which are related to the maximum 
oscillations in the $L_{\nu}/E_{\nu}$ distribution
through the correlation between $E_{\mu}$ and $E_{\nu}$ shown in 
Figure~14 in the preceding paper\cite{Konishi2}.
 However, it seems to be difficult to extract a pair of concrete 
values of $L_{\nu}$ and $E_{\nu}$ through the analysis of
the $L_{\nu}/E_{\mu}$ distribution.
 In Figure~\ref{figK044}, we make a comparison between
$L_{\nu}/E_{\nu}$ distribution and $L_{\nu}/E_{\mu}$ distribution where 
the correlation between $E_{\nu}$ and $E_{\mu}$ is shown in 
Figure~14 in the preceding paper\cite{Konishi2}.
 It is clear from the figure that the $L_{\nu}/E_{\nu}$ distribution 
demonstrates the maximum oscillation as already shown in 
Figures~\ref{figJ018} to \ref{figJ024} and the $L_{\nu}/E_{\mu}$ 
distribution also
demonstrates the maximum oscillation-like as already 
shown in Figure~\ref{figJ042} and \ref{figJ043}.
 In Figure~\ref{figK045}, we give the relation between 
$L_{\mu}/E_{\nu}$ distribution and 
$L_{\mu}/E_{\mu}$ distribution where the same correlation between 
$E_{\nu}$ and $E_{\mu}$ holds in the case of Figure~\ref{figK044}.
 It is also clear from the figures that both the distributions  
demonstrate neither the maximum oscillation nor the maximum 
oscillation-like, which is also clear from 
Figures~\ref{figJ029} to \ref{figJ033} and Figures~\ref{figJ038} to 
\ref{figJ039}. 
 Thus, it can be concluded from 
Figures~13 and 14 in the preceding paper\cite{Konishi2}
and Figure~\ref{figK044} and Figure~\ref{figK045} in the present paper
 that $L_{\nu}$ plays an essential role compared with others 
$L_{\mu}$, $E_{\nu}$ or $E_{\mu}$. 
In other words, it should be noticed that 
$L_{\nu}$ cannot be approximated by $L_{\mu}$, 
while $E_{\nu}$ can be obtained approximately from $E_{\mu}$ through 
some procedure.    
Also, such a serious discrepancy between 
$L_{\nu}$-$L_{\mu}$ relation and 
$E_{\nu}$-$E_{\mu}$ relation is shown in the comparison
 of Figure~\ref{figK044} with Figure~\ref{figK045}.  

\begin{figure}
\begin{center}
\resizebox{0.45\textwidth}{!}{%
  \includegraphics{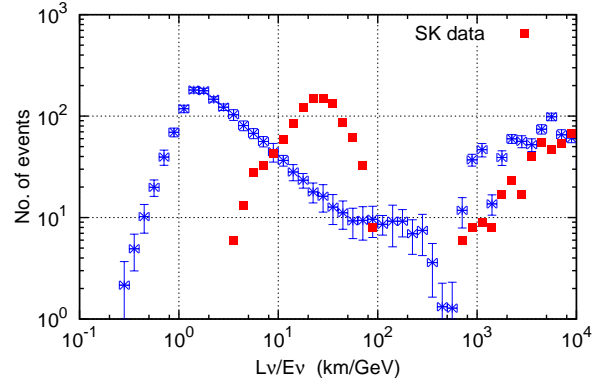}
}
\caption{The comparison of $L/E$ distribution for single-ring muon 
events due to QEL among {\it Fully Contained Events} with the 
corresponding one by the Super-Kamiokande
Experiment.}
\label{figK046}
\end{center}
\end{figure}

\section{Comparison of $L/E$ Distribution in the 
Super-Kamiokande Experiment with our Results}

In our classification, the $L/E$ distribution by Super-\\
Kamiokande Collaboration \cite {Ashie2}\cite {Ashie1}
should be compared directly with our $L_{\mu}/E_{\nu}$ distributiton. 
Taking account of their assertion
of existence of the maximum oscillation
we compare their results with our results on $L_{\nu}/E_{\nu}$
 in Figure~\ref{figK046}
\footnote
{We read out {\it Fully Contained Events} among total events
from Super-Kamiokande Collaboration \cite {Ashie2}\cite {Ashie1}.}. 
It is clear from the figure 
that there are two big differences between them. 

One is that we observe the first maximum oscillation 
($L_{\nu}/E_{\nu} = 515$ km/GeV under the SK 
oscillation parameters) sharply,
 while SK observe it in the wider range of  
$L_{\nu}/E_{\nu} = 100 \sim 800$ km/GeV.

\begin{figure}
\begin{center}
\resizebox{0.4\textwidth}{!}{%
  \includegraphics{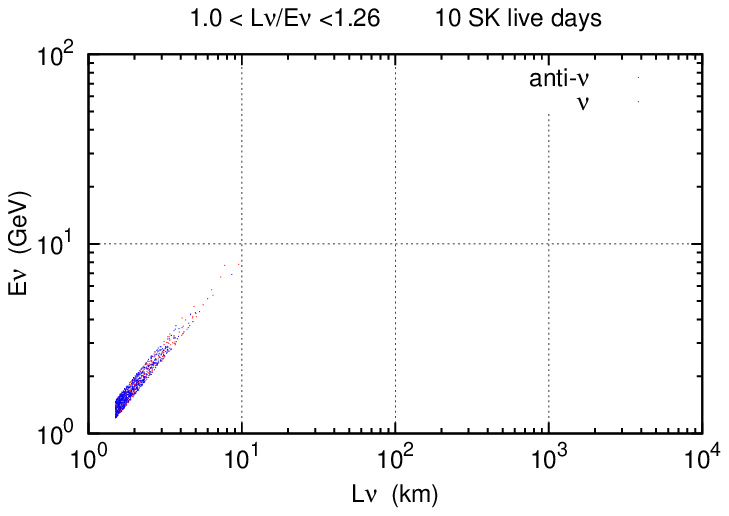}
}
\caption{Correlation diagram between $L_{\nu}$ and $E_{\nu}$ for 
$1.0<L_{\nu}/E_{\nu}<1.26 $ (km/GeV) which corresponds to the maximum frequency of the neutrino events for $L_{\nu}/E_{\nu}$
distribution in our computer numerical experiment for 14892 live days
 (10 SK live days).}
\label{figK047}
\resizebox{0.4\textwidth}{!}{%
  \includegraphics{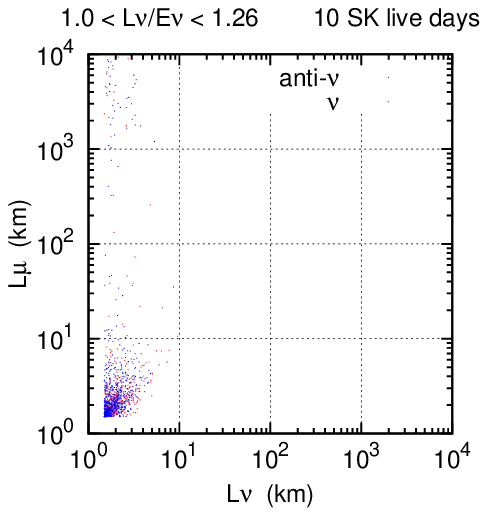}
}
\caption{Correlation diagram between $L_{\nu}$ and $L_{\mu}$ for 
$1.0<L_{\nu}/E_{\nu}<1.26 $ (km/GeV) 
under the neutrino oscillation parmeters obtained by 
Super-Kamiokande Collaboration
for 14892 live days (10 SK live days).}
\label{figK048}
\resizebox{0.4\textwidth}{!}{%
  \includegraphics{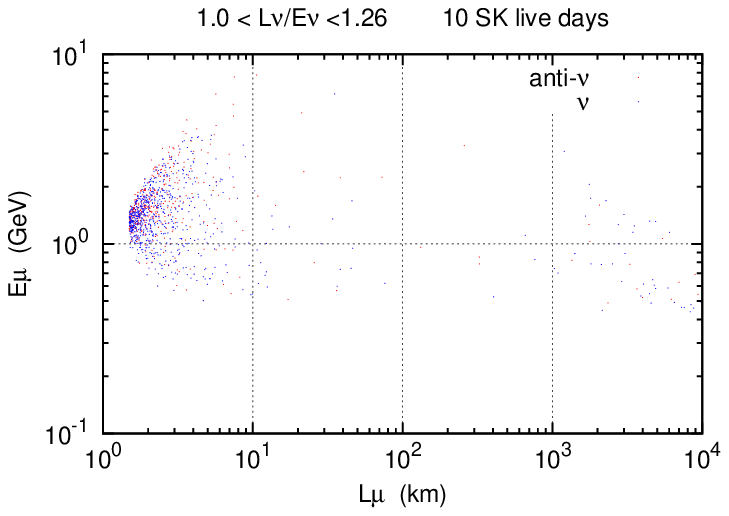}
}
\caption{Correlation diagram between $L_{\mu}$ and $E_{\mu}$ for 
$1.0<L_{\nu}/E_{\nu}<1.26 $ (km/GeV) which corresponds to the maximum frequency of the neutrino events for $L_{\nu}/E_{\nu}$
distribution in our computer numerical experiment for 14892 live days
(10 SK live days).}
\label{figK049}
\end{center}
\end{figure}

\begin{figure}
\begin{center}
\resizebox{0.4\textwidth}{!}{%
  \includegraphics{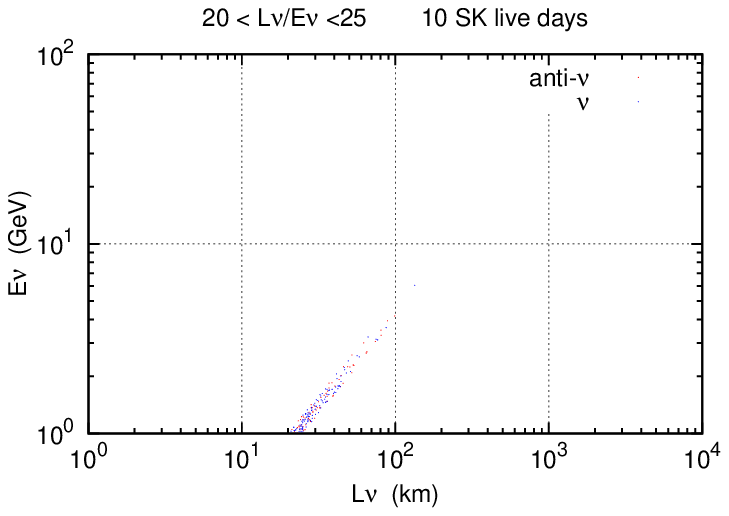}
}
\caption{Correlation diagram between $L_{\nu}$ and $E_{\nu}$ for 
$20<L_{\nu}/E_{\nu}<25 $ (km/GeV) which corresponds to the maximum frequency of the neutrino events for $L_{\mu}/E_{\nu}$
distribution in SK experiment for 14892 live days
(10 SK live days).}
\label{figK050}
\resizebox{0.4\textwidth}{!}{%
  \includegraphics{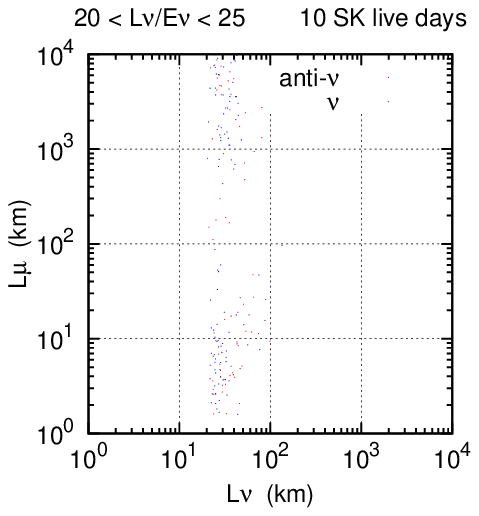}
}
\caption{Correlation diagram between $L_{\nu}$ and $L_{\mu}$ for 
$20<L_{\nu}/E_{\nu}<25 $ (km/GeV) 
under the neutrino oscillation parmeters obtained by 
Super-Kamiokande Collaboration
for 14892 live days (10 SK live days).}
\label{figK051}
\resizebox{0.4\textwidth}{!}{%
  \includegraphics{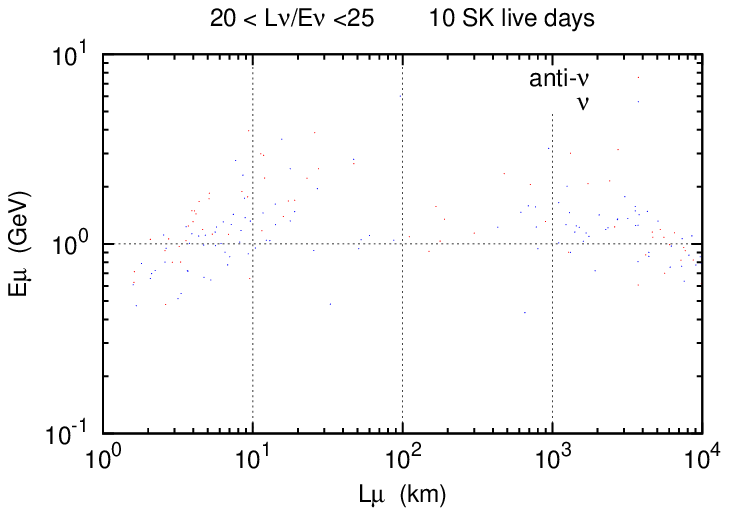}
}
\caption{Correlation diagram between $L_{\mu}$ and $E_{\mu}$ for 
$20<L_{\nu}/E_{\nu}<25 $ (km/GeV) which corresponds to the maximum frequency of the neutrino events for $L_{\nu}/E_{\nu}$
distribution in SK experiment for 14892 live days
(10 SK live days).}
\label{figK052}
\end{center}
\end{figure}

Such the lack of the neutrino events over the wide range may be due 
to their measurement of $L_{\mu}$, but not $L_{\nu}$,
because the given definite $L_{\nu}$ corresponds to 
$L_{\mu}$ over a wide range and vice versa 
(See also the correlation between $L_{\nu}$ and $L_{\mu}$ in
 Figure~\ref{figK048} and \ref{figK051})


 The other is that there is big difference between them as for the 
position which give the maximum frequency for the events concerned.
Here, we do not mention to the existence of the maximum oscillation 
which is derived from the measurement of $L_{\mu}$ utilized in 
Super-Kamiokande Collaboration,
 because one cannot observe the maximum oscillation, if we 
utilize $L_{\mu}$ (see Figures~\ref{figJ028} to \ref{figJ033}). 
 Conseqently, 
we examine the second point as for the maximum frequency for the 
events concerned. Our computer numerical experiment gives the maximum 
frequency for interval $1.0<L_{\nu}/E_{\nu}<1.26 $ (km/GeV)
 as shown in Figure~\ref{figK046}.

 In Figure~\ref{figK047}, we give the correlation between
$L_{\nu}$ and $E_{\nu}$ for interval $1.0<L_{\nu}/E_{\nu}<1.26 $ (km/GeV). 
It is clear from the figure that the larger part of the incident neutrino 
events is occupied by the vertically downward ones and the smaller part 
is occupied by the horizontally downward neutrino events. 
 This is quite reasonable, because 
more intensive downward flux contribute to the maximum frequency for the 
events concerned, compared with weaker upward flux under the 
Super-Kamokande neutrino oscillation parameters. 

\begin{figure}
\begin{center}

\resizebox{0.4\textwidth}{!}{%
  \includegraphics{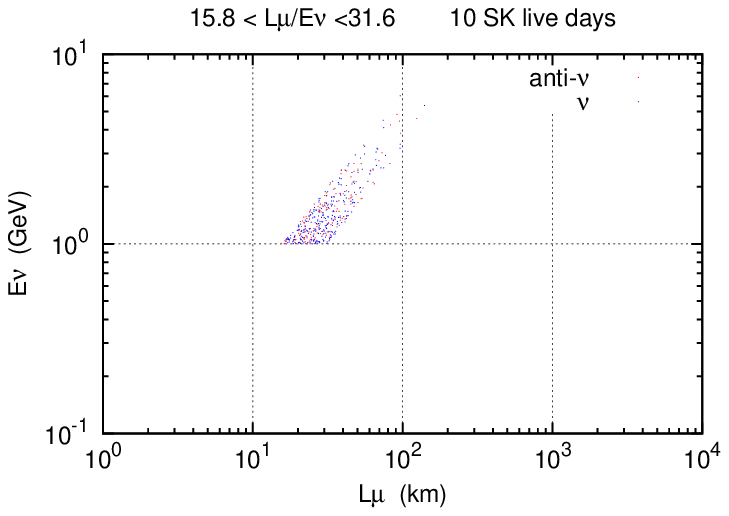}
}
\caption{Correlation diagram between $L_{\mu}$ and $E_{\nu}$ for 
$15.8<L_{\mu}/E_{\nu}<31.6 $ (km/GeV) which correspond to the maximum 
frequency of the neutrino events for $L_{\mu}/E_{\nu}$
distribution in SK experiment for 14892 live days
(10 SK live days).}
\label{figK053}
\resizebox{0.4\textwidth}{!}{%
  \includegraphics{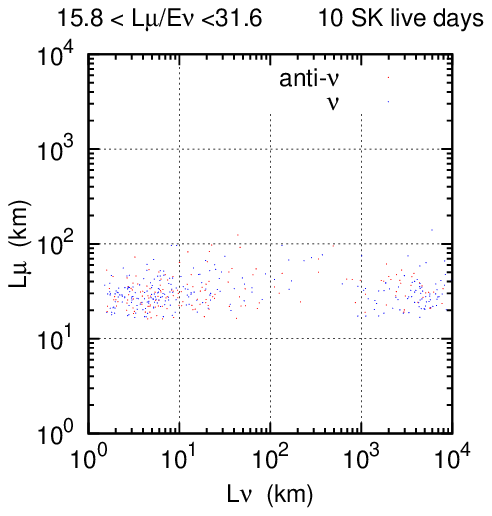}
}
\caption{Correlation diagram between $L_{\nu}$ and $L_{\mu}$ for 
$15.8<L_{\mu}/E_{\nu}<31.6 $ (km/GeV) 
under the neutrino oscillation parmeters obtained by 
Super-Kamiokande Collaboration
for 14892 live days (10 SK live days).}
\label{figK054}
\resizebox{0.4\textwidth}{!}{%
  \includegraphics{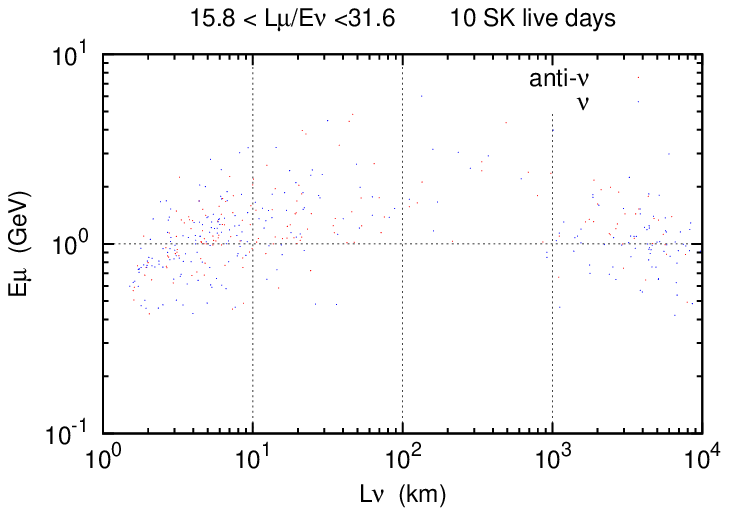}
}
\caption{Correlation diagram between $L_{\nu}$ and $E_{\mu}$ for 
$15.8<L_{\mu}/E_{\nu}<31.6 $ (km/GeV) which correspond to the maximum 
frequency of the neutrino events for $L_{\mu}/E_{\nu}$
distribution in SK experiment for 14892 live days
(10 SK live days).}
\label{figK055}
\end{center}
\end{figure}
In Figure~\ref{figK048}, we give the correlation diagram between 
$L_{\nu}$ and $L_{\mu}$ for the same intervals as in Figure~\ref{figK047}.
It is clear from Figure~\ref{figK048} that the majority  of the events is 
concentrated into the squared regions with $L_{\nu}<10$~km and 
$L_{\mu}<10$~km. This denotes that the downward incident 
neutrinos produce muons toward the forward direction
with either smaller or larger angles 
and only the smaller part of the downward incident neutrino events 
produce the upward muons due to backscattering 
(1000 to 10,000 km in $L_{\mu}$)
 as well as the azimuthal angle effect in QEL.
 In Figure~\ref{figK049}, we give the correlation diagram between 
$L_{\mu}$ and $E_{\mu}$ for the same intervals as in Figure~\ref{figK047}.
   It is clear from this figure that the produced muons with 
higher energies are ejected toward the forward and vertical-like
directions, while the produced muon with lower energies may be ejected 
toward the backward or holizontal-like direction. 
Namely, it is concluded from 
Figures~\ref{figK047}, \ref{figK048} and \ref{figK049} that the vertically 
downward neutrino events can contribute to the maximum frequency, because
they are free from neutrino oscillation. It is further noted that the 
direction of the produced muon does not coincide with the original direction of the neutrino. 

 Now, we examine $L$-$E$ relation at the position for our computer 
numerical expermiment  where Super-Kamiokande Collaboration give the 
maximum frequency for the events ($20<L_{\nu}/E_{\nu}<25 $ (km/GeV)).
 In Figure~\ref{figK050}, we show the correlation between 
$L_{\nu}$ and $E_{\nu}$ for interval $20<L_{\nu}/E_{\nu}<25 $ (km/GeV).

It is clear from Figure~\ref{figK050} that 
$L_{\nu}$ distribute over 27 $\sim$ 120 km, corresponding to 
$cos\theta_{\nu} =$ - 0.1 $\sim$ 0, which denotes the horizontal-like 
downward neutrino events. 
 The frequency of the horizontal-like downward neutrino events in
 Figure~\ref{figK050} are pretty smaller than that of the 
vertical-like downward neutrino events in Figure~\ref{figK047}
due to smaller solid angles.  
 In Figure~\ref{figK051}, we give the correlation diagram between
$L_{\nu}$ and $L_{\mu}$ for $20<L_{\nu}/E_{\nu}<25 $ (km/GeV).
 It is impressive from the figure that $L_{\mu}$
distribute over four orders of magnitude (2~km to $1.2\times 10^4$ km),
 while $L_{\nu}$ cover within one order of
magnitude (20 $\sim$ 120 km).
 This fact denotes that the effect of the azimuthal angles in QEL is 
pretty strong even in the horizontal-like downward neutrino events 
in which the produced muons are apparently judged to
come from the upward direction 
(see Figure~3-c and Figures~8 to 10 in the preceding 
paper\cite{Konishi2}).

In Figure~\ref{figK052}, we give the correlation diagram between 
$L_{\mu}$ and $E_{\mu}$ for the same intervals as in Figure~\ref{figK050}.
If we compare Figure~\ref{figK050} with Figure~\ref{figK052},
 then we can find the following interesting situation.
As it is clearly understandable from Figure~\ref{figK050},
 horizontal-like downward neutrinos 
 produce the muons in the three 
different regions, namely, vertical-like downward muons, 
horizontal-like downward muons and upward muons.
From horizontal-like downward neutrinos with rather low energies,
the vertically 
downward muons are ejected with rather large scattering angles.  
On the other hand, the horizontal-like 
downward muons are ejected with rather small angles whose energies are 
close to the incident neutrinos energies.
Furthermore, the upward muons are 
produced either due to backscattering or due to the azimuthal effect in 
QEL for horizontal-like incident neutrinos 
(see Figures~8 and 9 in the preceding paper\cite{Konishi2}). 
 Thus, from the comparison of 
Figures~\ref{figK047}, \ref{figK048} and \ref{figK049} with 
Figures~\ref{figK050}, \ref{figK051} and \ref{figK052},
 it is reasonable for the the maximun frequency of the 
$L_{\nu}/E_{\nu}$ events to occur for 
$1.0<L_{\nu}/E_{\nu}<1.26 $ (km/GeV),
 and not to occur for $20<L_{\nu}/E_{\nu}<25 $ (km/GeV) 
 where Super-Kamiokande Collaboration "assert".

Finally, we examine the correlation between
 $L_{\mu}$ and $E_{\nu}$
for $15.8<L_{\mu}/E_{\nu}<31.6 $ (km/GeV)
 where Super-Kamiokande Collaboration give
the maximum frequency of $L/E$ neutrino events
as shown in Figure~\ref{figK046}.   
Although we compare their frequency with that of our 
$L_{\nu}/E_{\nu}$ in Figure~\ref{figK046},  we can  compare their frequency with that of our $L_{\mu}/E_{\nu}$ in Figure~\ref{figK045},
 which shows big difference between them. 
In Figure~\ref{figK053}, we give the correlation diagram between 
 $L_{\mu}$ and $E_{\nu}$ for $15.8<L_{\mu}/E_{\nu}<31.6 $ (km/GeV).
  In Figures~\ref{figK054} and \ref{figK055},
 we give the corresponding correlation diagrams 
between $L_{\nu}$ and $L_{\mu}$, and $L_{\nu}$ and $E_{\mu}$,
respectively.

It is clear from Figure~\ref{figK053} that Super-Kamiokande Collaboration 
measure the vertical-like downward muons. 
It is also clear from Figures~\ref{figK054} and \ref{figK055}
that these vertical-like downward muon are produced by the incident 
neutrinos whose $L_{\nu}$ are distributed over four orders of magnitude.
 These incident neutrinos are classified into two parts.
 One is the downward incident neutrinos ($1.0<L_{\nu}<100$ km)
 and the other ($L_{\nu}>100$ km) is the upward incident neutrinos.
The majority of the incident neutrino is occupied by the 
vertical-like downward.
 However, the frequency of the upward neutrinos is 
in the same order of the magnitude as the horizontal-like downward.
 The upward incident neutrinos may produce downward muons due to 
both backscattering and the azithumal angle effect in QEL.
  At any rate, for the measured muons in the case of the maximun 
frequency of the events, $L_{\nu}$ of the corresponding incident 
neutrinos distribute over four orders of magnitude.
 Shortly speaking, for the maximum frequency of the neutrino events
 $15.8<L_{\mu}/E_{\nu}<31.6 $ (km/GeV), the magnitude of 
the $L_{\mu}$ of the produced muons lie within one order of 
magnitude (see Figure~\ref{figK053}), although the $L_{\nu}$ of the 
incident neutrinos which produce these muons distribute over four orders
 of magnitude. In other words, 
it is concluded that Super-Kamiokande Collaboration do not measure the 
definite direction of the incident neutrinos as far as they measure 
$L_{\mu}$. 
It is furthermore noticed from the comparison of 
Figure~\ref{figK055} with Figure~\ref{figJ043}
 that Figure~\ref{figK055} is obtained from 
Figure~\ref{figJ043} by cutting  off the stripe of 
$15.8<L_{\mu}/E_{\nu}<31.6 $ (km/GeV).
 Therefore ,we can recognize the vacant region of the neutrino events
 faintly in the part between 100 and 1000 (km/GeV) in 
Figure~\ref{figK055} which is clearly shown in 
Figure~\ref{figJ043}. The vacant region of the events 
shows indication of neutrino oscillation.  

The summary on Figures from \ref{figK047} to \ref{figK055} are as follows;
 Figures from  \ref{figK047} to \ref{figK049} represent the mutual relations among $L_{\nu}$, $L_{\mu}$, $E_{\nu}$ and $E_{\mu}$
near  {\it our} maximum frequency of $L_{\nu}/E_{\nu}$ distribution. 
Here, all the incident neutrinos are occupied by the downward 
vertical-like neutrinos, while the majority of the emitted muons is 
occupied by the downward muon and the minority is 
occupied by the upward muon. 
 Figures from \ref{figK050} to \ref{figK052} represent the similar  
mutual relations for {\it our} $L_{\nu}/E_{\nu}$ distribution which 
correspond to the near the maximum frequency of
$L_{\mu}/E_{\nu}$ distribution obtained by Super-Kamiokande 
Collaboration. Here, almost the incident neutrinos are occupied  
by the downward holizontal-like neutrinos, while about the half of 
emitted muons is recognized as the downward muon and the other half is 
done as the upward muon. 
 Figures from \ref{figK053} to \ref{figK055} represent the mutual 
similar relations, assuming  the numerical 
values of the maximum frequency of 
$L_{\mu}/E_{\nu}$ distribution obtained by 
Super-Kamiokande Collaboration.
 Here, the majority of the emitted muons is 
occupied by the horizontal like muons, 
while their parent neutrinos come from both 
the downward neutrinos and the upward ones.
  The common characteristics through
 Figures from \ref{figK047} to \ref{figK055} 
is that for given definite $L_{\nu}$($L_{\mu}$)  we  find
 $L_{\mu}$($L_{\nu}$) which distribute over the four order of magnitudes.

\section{Conclusion}
The assumption made by Super-Kamiokande Collaboration that the 
direction of the reconstructed lepton approximately represents the 
direction of the original neutrino does not hold even approximately
\cite{Konishi1}.
 This is logically equivalent to the statement that 
$L_{\nu}$ cannot be replaced by $L_{\mu}$ even if approximately.
 This is really  clarified in 
Figures~12 and 13 in the preceding paper\cite{Konishi2}.   

 Although the derivation of $E_{\nu}$ from $E_{\mu}$ (Eq.(6) of
the preceding paper\cite{Konishi2}) is 
theoretically, irrelevant to the stochastic plobrem,
 because of the neglect of the stochastic 
character in physical processes concerned, such the approximation does 
not induce so practically
serious error compared with the assumption of 
 $L_{\nu} \approx L_{\mu}$. 
 As clarified in Figures~\ref{figJ018} to \ref{figJ026},
 the maximum oscillation in $L/E$ 
analysis can be observed only in the $L_{\nu}/E_{\nu}$
  distribution and it is quite natural by the definition of the 
probability for a given favor whose argument is $L_{\nu}/E_{\nu}$
(Eq.(1)).
  As clarified in Figures~\ref{figJ029} to \ref{figJ033} and 
Figures~\ref{figJ038} to ~\ref{figJ040} the maximum oscillation 
for the presence of neutrino oscillation cannot be observed from both 
$L_{\mu}/E_{\mu}$ and $L_{\mu}/E_{\nu}$.
The relation between 
 $L_{\nu}$ and  $L_{\mu}$ is too complicate to 
extract
similar expression to Eq.(6) of the preceding paper\cite{Konishi2}
for the argument on 
$L_{\mu}/E_{\mu}$ and $L_{\mu}/E_{\nu}$ . 
Similarly in the case of argument of $L_{\nu}/E_{\nu}$,
 we can indicate something like the maximum oscillation in 
$L_{\nu}/E_{\mu}$ distribution which are shown in 
Figures~\ref{figJ041}41 and \ref{figJ042}.
 The situation is derived from the fact that what plays a decisive role 
in $L/E$ analysis is $L_{\nu}$, but not $E_{\nu}$, which are clearly 
shown by comparing 
Figures~12 and 13 with Figure~14 in the preceding paper\cite{Konishi2}.

As for $L/E$ distribution obtained by Super-Kamiokande Collaboration,
 we  definitely indicate that the maximum oscillation cannot be 
observed through the measurement of $L_{\mu}$.
 Consequently, we cannot observe the maximum oscillation in 
$L/E$ analysis which is carried out in Super-Kamiokande Collaboration. 
Furthermore, one cannot find the maximum  frequency of $L/E$ events at 
the position where Super-Kamiokande Collaboration observe,
 even if one can observe $L_{\nu}$.   

In conclusion, the maximum oscillation in $L/E$ analysis can be observed 
only in $L_{\nu}/E_{\nu}$, but not in any other combinations 
of $L$ with $E$. 
However, $L_{\nu}$ is physically unobservable quantities and 
it cannot be approximated by $L_{\mu}$, 
because the assumption between $L_{\nu}$ and $L_{\mu}$, does not hold
even if statistically. 
Consequently, it should be concluded that Super-Kamiokande cannot observe 
the maximum oscillation in their $L_{\mu}/E_{\nu,SK}$ analysis. 

Finally, our conclusion that $L_{\nu}$ cannot be approximate 
by $L_{\mu}$ is logically equivalent to the statement that 
$cos\theta_{\nu}$ cannot be approximated by 
$cos\theta_{\mu}$,
 where $cos\theta_{\nu}$ denotes cosine of the zenith angle of the 
incident neutrino and $cos\theta_{\mu}$ denotes
 that of the produced muon, respectively \cite {Konishi1}.
 In Super-Kamiokande Collaboration, they approximate
$cos\theta_{\nu}$ as $cos\theta_{\mu}$
 ( See the reproduction of their statements in the 2 page 
in the present paper ). 
The analysis of the zenith angle distribution of the atmospheric neutrino 
events by Super-Kamiokande Collaboration will be re-examined in our 
subsequent papers.

%


\begin{thebibliography}{}
%
%
  \bibitem{Konishi2} Konishi,E {\it et al.}, arXiv hep-ex/1007.3812v1
 \bibitem{Ashie2} Ashie,Y. {\it et al.}, Phys. Rev. D {\bf 71} (2005) 112005.
  \bibitem{Honda} Honda, M., {\it et al.}, \  Phys.\ Rev. D {\bf 52} (1996) 4985.\\
 Honda, M., {\it et al.}, \  Phys.\ Rev. D {\bf 70} (2004)043008-1. 
  \bibitem{Ashie1} Ashie,Y {\it et al.}, Phys.Rev.Lett.{\bf93}
(2004)101801-1.
  \bibitem{Konishi1} Konishi,E {\it et al.}, arXiv hep-ex/0808.0664v2


\end{thebibliography}
\end{document}